# Exploration of Enterprise Big Data Microservice Architecture Based on Domain-Driven Design (DDD)

Yiru Zhang
*Cornell Tech*
2 W Loop Rd, New York, NY 10044, USA
yiruz315@163.com

*Abstract*—With the rapid advancement of digitization and intelligence, enterprise big data processing platforms have become increasingly important in data management. However, traditional monolithic architectures, due to their high coupling, are unable to cope with increasingly complex demands in the face of business expansion and increased data volume, resulting in limited platform scalability and decreased data collection efficiency. This article proposes a solution for enterprise big data processing platform based on microservice architecture, based on the concept of Domain Driven Design (DDD). Through in-depth analysis of business requirements, the functional and non functional requirements of the platform in various scenarios were determined, and the DDD method was used to decompose the core business logic into independent microservice modules, enabling data collection, parsing, cleaning, and visualization functions to be independently developed, deployed, and upgraded, thereby improving the flexibility and scalability of the system. This article also designs an automated data collection process based on microservices and proposes an improved dynamic scheduling algorithm to efficiently allocate data collection tasks to Docker nodes, and monitor the collection progress and service status in real time to ensure the accuracy and efficiency of data collection. Through the implementation and testing of the platform, it has been verified that the enterprise big data processing platform based on microservice architecture has significantly improved scalability, data quality, and collection efficiency.

## I. Introduction

With the rapid development of digitization and intelligence, the big data industry has made rapid progress. Emerging technologies such as microservices, cloud computing, and artificial intelligence have been widely applied, bringing profound changes to society, promoting the formation of a big data processing ecosystem, and providing more diversified services. However, the increasingly complex business requirements, continuous increase in new modules and data flows, pose severe challenges to the reliability, scalability, and maintainability of big data processing platforms. With the progress of society, the demand for big data processing platforms from enterprises is rapidly increasing. Traditional architectures are no longer able to effectively support the expanding business functions, making the rationality and standardization of platform architecture increasingly important. Currently, the main problem with enterprise big data processing platforms is that their single architecture is difficult to adapt to diverse needs and multifunctional expansion scenarios. The traditional integrated data governance model is no longer sufficient to meet the business needs of the big data era. The highly coupled code structure increases the difficulty for developers in troubleshooting and solving problems, thereby affecting user experience and hindering the ability of enterprises to conduct real-time data analysis. Based on this, this article proposes to transform the enterprise big data processing platform through microservice architecture to enhance the platform's service capabilities, improve system availability, reduce code coupling, improve scalability and maintenance efficiency, thereby reducing future development and maintenance costs. This article also designs a new ETL process, including data collection, Kafka message passing, data parsing, Spark processing, data cleaning, and visualization, combined with automatic monitoring and restart mechanisms, greatly improving the efficiency and quality of data collection. By analyzing the shortcomings of traditional monolithic architecture, this article explores the design and implementation of an enterprise big data processing platform based on microservice architecture. The entire process strictly follows the basic principles of software engineering, covering various aspects such as requirement analysis, system design, development and implementation, and testing.

## II. Related Research

### A  DDD Domain Driven Design

Design methods like DDD are powerful and effective tools for refactoring and solving software problems. It provides a standardized framework for software development and restructuring, despite the complexity and cognitive burden involved in its application. In the study of O Zkan et al., they reconstructed existing software systems based on the DDD principle[1]. Design methods like DDD are powerful and effective tools for refactoring and solving software problems. It provides a standardized framework for software development and restructuring, despite the complexity and cognitive burden involved in its application. In the study of O Zkan et al., they reconstructed existing software systems based on the DDD principle. Y Kong et al. proposed a new sparse classification method called Data Driven Dictionary Based Sparse Classification (DDD-SC)[2]. This method does not require the detection of low-frequency features or the construction of explicit classifiers, and can effectively identify faults in planetary bearings. Domain driven design (DDD) is a key design method in microservice architecture (MSA), which effectively supports the design of architecture components during model creation and implementation by adopting a series of standard patterns[3]. S. Giallorenzo et al. explored advanced technologies for microservice design and implementation in their research[4]. They propose that developers can build domain models in microservice architectures through Domain Driven Design (DDD) and automatically translate them into programming languages that support service orientation, which natively support the behavioral implementation of each microservice.

### B  Scheduling algorithm

K. Kia et al. introduced a dynamic heuristic scheduling algorithm called HDSAP in their study, which utilizes shared memory multi-core processors to address the side effects generated under uncertain workloads[5]. In order to ensure the reliability of the system, they adopted N-mode redundancy technology and considered the variation of kernel soft error rate to prevent inaccurate estimation of reliability. D Liu et al. proposed a novel multi strategy discrete constrained differential evolution algorithm called MSDCDE in their research[6]. This algorithm combines discrete variable crossover strategy, constraint handling method, population restart mechanism, and left shift local strategy, aiming to effectively avoid falling into local optima and achieve global optimal solutions, thereby improving search performance and rescue effectiveness.

### C  Data Warehousing and Data Mining

The regression model proposed by AH Carlson[7] adopts the two-step consensus estimator of the Heckman method. This method is similar to the traditional two-step consistent Heckman estimation, but it allows the existence of heteroscedasticity in the first step and is more general in the setting of the control function. In their study[8], AC Jaures et al. analyzed the local seasonal climate prediction (ISCF) of Benin, combined with the travel cost method, descriptive statistical methods, and the two-step Heckman model, to evaluate the application and economic value of this prediction. Krishnan et al. [9] also adopted Heckman's two-step method (1979) in their research. The results showed that the European Central Bank (ECB) has a positive impact on corporate foreign direct investment (OFDI), and pointed out that enterprises with high leverage levels and ECB participation tend to exhibit a higher scale of OFDI. Furthermore, MAJi-Liang et al. (bbb) used the Heckman two-step model to identify the factors influencing the decision-making of commercial leguminous crop cultivation households[10]. They also applied the endogenous processing regression (ETR) method to assess how commercial leguminous crop cultivation affects family economic welfare.

## III.   METHOD

### A  Design of an Enterprise Big Data Automatic Processing Platform Based on Microservices

The overall process of the enterprise big data automatic processing platform discussed in this article covers data collection, parsing, cleaning, visualization, and backend management. Users can manage information through the backend center. According to this process, the platform is divided into six core subdomains, including user center, collection center, parsing center, data cleaning, data visualization, and backend center. Each subdomain is transformed into a microservice, which includes data collection microservice, data parsing microservice, data cleaning microservice, data visualization microservice, backend center microservice, and user center microservice. This modular design enhances the scalability and flexibility of the system, allowing each microservice to be independently updated and maintained, optimizing overall functionality and integration effectiveness. The data is transmitted through Kafka message queues and message types are identified using different topics, such as SEARCHPAGE, ITEMPAGE, and COMMENTPAGE. The parsing program extracts data from the message queue. Data cleaning is performed through Spark Streaming for streaming processing, dividing real-time data into multiple micro batches and processing each batch based on time nodes to achieve fast response and real-time data processing. The complete process of ETL is shown in Figure 1.

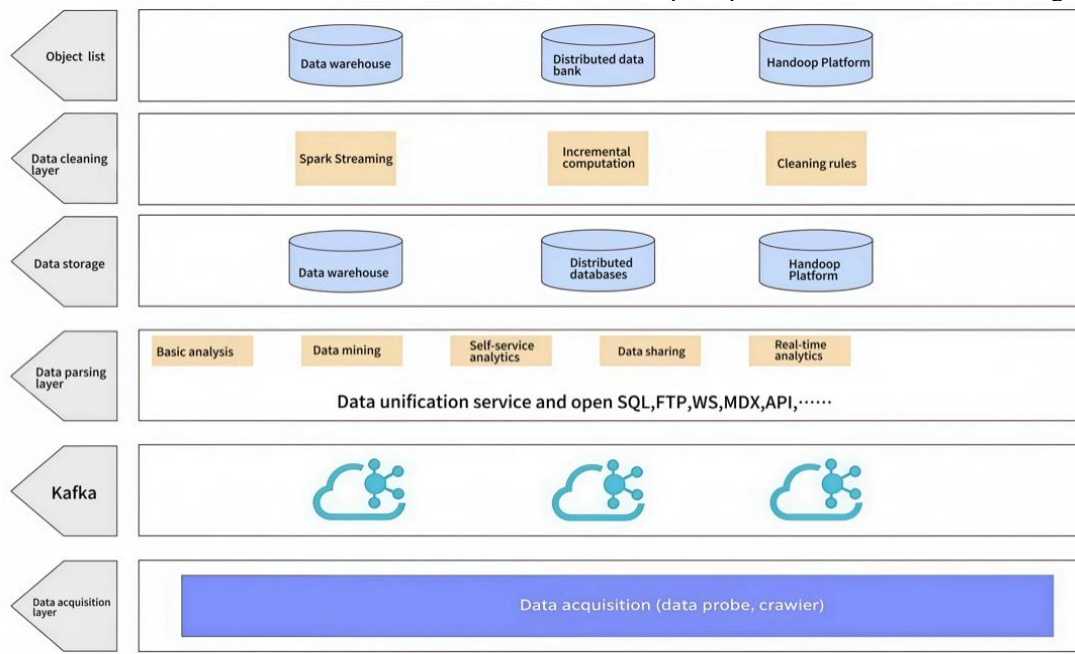

Figure 1. ETL overall process diagram

In Domain Driven Design (DDD), the concept of aggregation is used to integrate related domain objects into an independent unit for easy management and maintenance. In the enterprise big data automatic processing platform, this means aggregating functional modules such as data collection, parsing, cleaning, and visualization into a unified data processing unit. Each aggregation consists of a root entity and multiple related entities and value objects, where the root entity is responsible for coordinating and managing other components. The database design of microservice architecture is equally crucial, emphasizing the independence of the database, that is, each microservice should have an independent database. Only the microservices corresponding to the database can be directly accessed, while other microservices need to operate through interfaces, which can reduce direct operations, alleviate database burden, and prevent unauthorized access. Meanwhile, data sharing between services is achieved through static tables and master-slave backup tables. Each microservice maintains a one-to-one relationship with its database, which not only optimizes service interfaces, improves data security and stability, but also enhances system scalability and maintainability.

*B   Design of Automatic Monitoring and Data Collection*

This article explores how to introduce scheduling algorithms into automatic monitoring and collection systems to achieve automation of data collection. By studying common scheduling algorithms, we analyzed their advantages and disadvantages, and proposed an improved DSOM scheduling algorithm based on genetic algorithm, specifically optimized for data collection scenarios. This article provides a detailed description of the specific steps and process of the DSOM algorithm, and demonstrates its advantages in multiple performance metrics through experimental results. Meanwhile, we compared the DSOM algorithm with other scheduling methods and verified its significant effectiveness in improving task scheduling efficiency, optimizing load balancing, and reducing energy consumption. Ultimately, this optimized scheduling algorithm was applied to automatic monitoring and collection services, effectively improving collection efficiency. In addition, we also discussed how to schedule resources in Docker based microservice architecture to optimize resource allocation for data collection services. Although genetic algorithms can handle complex optimization problems due to their universal applicability and powerful capabilities, their effectiveness is affected by parameter settings. Therefore, this article introduces a container fitness evaluation function that adapts to the characteristics of data collection microservices, and combines the DSOM scheduling algorithm to further optimize microservice resource scheduling.

When applying the DSOM scheduling algorithm for microservice resource scheduling, the first step is to encode the problem and initialize the population. Next, the scheduling effectiveness of each individual is evaluated using the fitness function (see formula 1) to determine the rationality of resource allocation. In the formula, D represents the resource utilization of the physical host, Z/U is the load balancing rate, Z is the host load value, U is the average load value, and N represents the communication cost. The fitness function is expressed as:

$$Fitness = 1000\left(\alpha \frac{1}{D} + \beta \left(\frac{Z}{U}\right) + \lambda \frac{1}{1000N}\right) \quad (1)$$

Next, the selection, crossover, and mutation operations of the DSOM scheduling algorithm are used to generate the first generation offspring population. The selection operation adopts the tournament selection method to compare individual fitness, and the better one enters the next generation. Cross operation includes gene exchange, deletion, and supplementation, all of which are adjusted according to fitness functions. Mutation operation identifies the individual with the lowest fitness, removes its corresponding physical host, and reallocates the original microservice. Starting from the second generation, merge the newly generated offspring with the parent population to form a new parent population. Then repeat the selection, crossover, and mutation operations until the termination conditions are met. Finally, by combining the automatic monitoring and collection process with the DSOM scheduling algorithm, the goal of automated and efficient data collection was achieved.

*C   Microservice Governance Design*

The platform adopts SpringCloudAlibaba components and Docker for service deployment, and implements unified management of Docker through Kubernetes. The service configuration is handled by Nacos, the service gateway is completed by the Gateway component, and the registration and discovery of services rely on Kubernetes' Service mechanism. All application deployments are orchestrated through the combination of Docker and Kubernetes, which not only improves system management efficiency but also simplifies service management. Under the microservice architecture, although the splitting of services improves system flexibility and iteration speed, it also brings challenges in managing complexity. To address these challenges, we have chosen Kubernetes' Service mechanism, which outperforms traditional Eureka and Nacos in performance and is more suitable for the governance needs of microservices. Kubernetes' Service provides a unified access point for container applications with similar functionality, and distributes requests across multiple backend containers through load balancing, thereby optimizing service performance and stability.The working mechanism of NodePort type Service is shown in Figure 2.

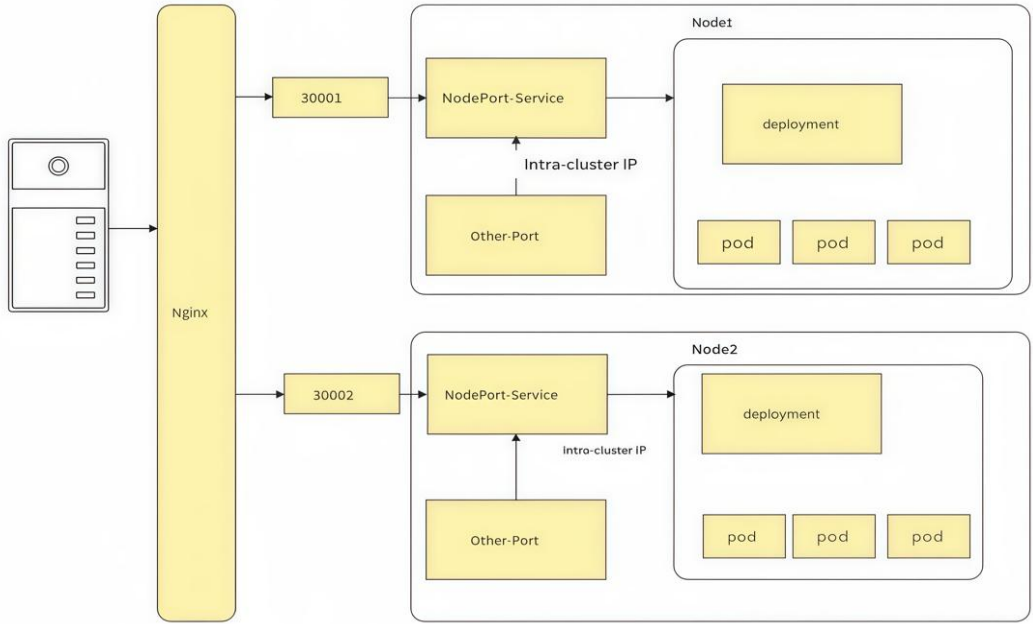

Figure 2. Service operation mechanism of enterprise big data automatic processing platform

## IV. RESULTS AND DISCUSSION

### A  Automatic monitoring, collection, and testing

To evaluate the effectiveness of the DSOM scheduling algorithm in resource allocation, three different scale experimental environments will be established. These environments include a small (ten), medium (one hundred), and large (one thousand) microservice containers, designed to simulate the actual performance of enterprise big data automated processing platforms at various deployment scales. Table 1 shows the configuration details of microservice containers and corresponding physical machines in various scale environments, which ensures the applicability and efficiency of the comprehensive testing algorithm in different scale scenarios.

TABLE 1. CONTAINER AND PHYSICAL MACHINE SCALE CONFIGURATION

| Configuration Name | Number of microservice containers | Number of physical machines |
|---|---|---|
| S1 | 80 | 8 |
| S2 | 358 | 35 |
| S3 | 1125 | 112 |

Due to CloudSim's three-layer architecture of "container virtual machine physical machine", the "container physical machine" architecture is sufficient to meet the requirements in enterprise big data automatic processing platforms based on microservices. This simplified architecture can improve the response speed of microservices. In CloudSim simulation, we set the resource levels of physical and virtual machines to be the same in order to simulate the actual deployment of microservice platforms. To evaluate the performance of the DSOM algorithm, we conducted comparative tests with maximum utilization first, first come first served, and polling algorithms. In the experiment, the algorithm parameters were set to $\{\alpha=0.2, \beta=0.5, \gamma=0.3\}$ to adapt to different resource scheduling scenarios. The test results show that as the scale of microservice containers increases, the usage of physical machines also increases, but the DSOM algorithm always maintains the lowest physical machine usage. Meanwhile, Figures 3 and 4 demonstrate that the DSOM algorithm outperforms other algorithms in terms of resource utilization, providing higher resource utilization efficiency in various scale scenarios.

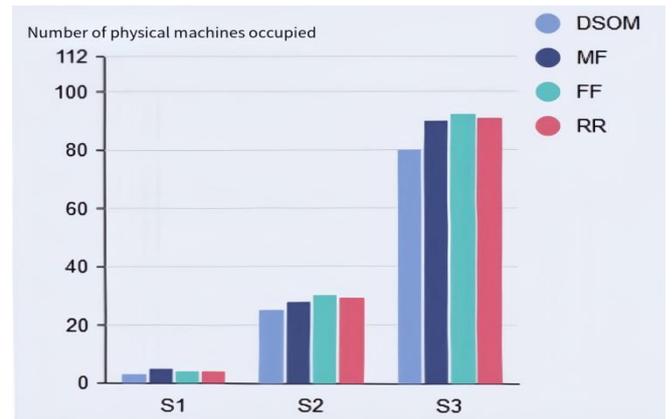

Figure 3. Comparison of the number of physical machines occupied by four algorithms at different scales

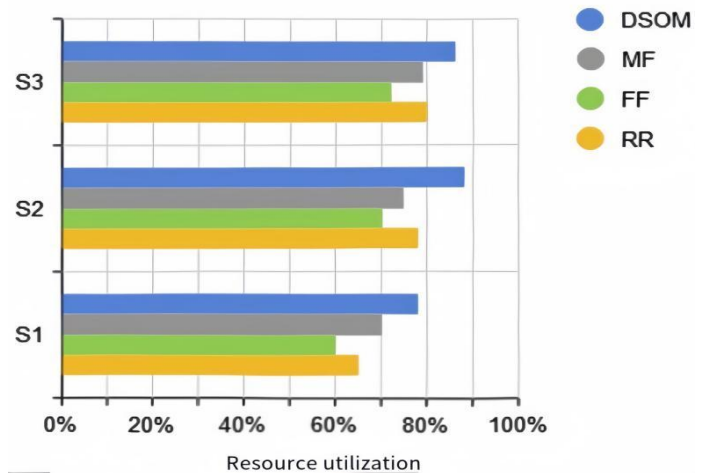

Figure 4. Comparison of resource utilization rates of four algorithms at different scales

As shown in Figure 5, with the expansion of microservice deployment scale, communication costs correspondingly increase. However, compared to the other three algorithms, the DSOM algorithm has lower communication costs and growth rates at the same deployment scale. This is because the DSOM algorithm takes into account the communication costs between microservices during deployment, prioritizing the deployment of microservices with lower communication costs on the same physical machine, thereby effectively reducing the overall communication overhead of the system.

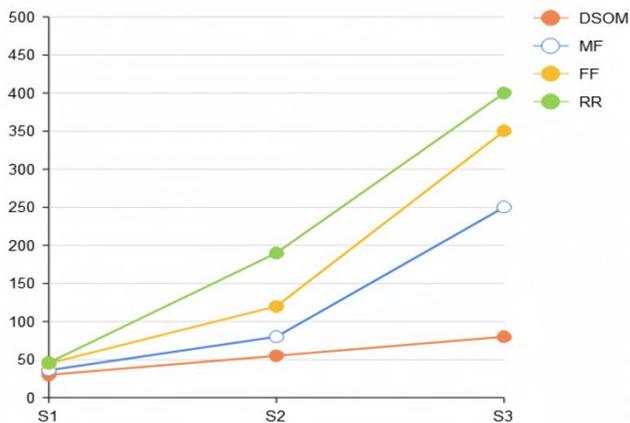

Figure 5. Comparison of communication costs between microservice containers using four algorithms at different scales

### B Non functional testing of microservice governance

In order to evaluate the reliability of the platform, this article uses the Jmeter pressure testing tool for performance testing, with concurrent thread counts ranging from 1000 to 5000. Figure 6 shows the throughput of the platform under different concurrency conditions, while Figure 7 shows that when the number of concurrent threads reaches 5000, the CPU and memory utilization rates have almost reached their maximum. Therefore, it can be concluded that the maximum concurrent support of this platform is close to 5000, demonstrating high reliability.

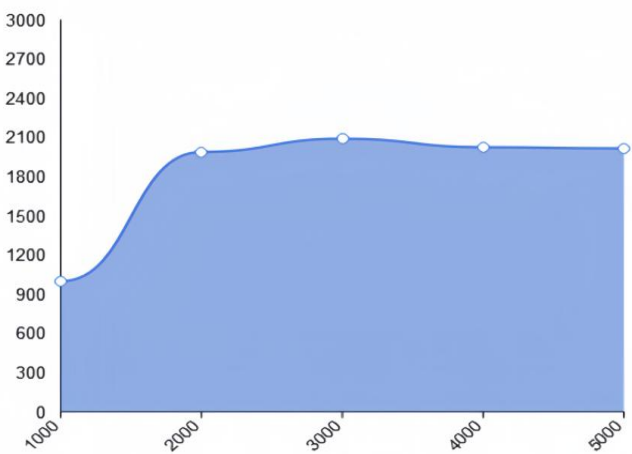

Figure 6. Throughput Trend Chart of Enterprise Big Data Automatic Processing Platform Based on Microservices

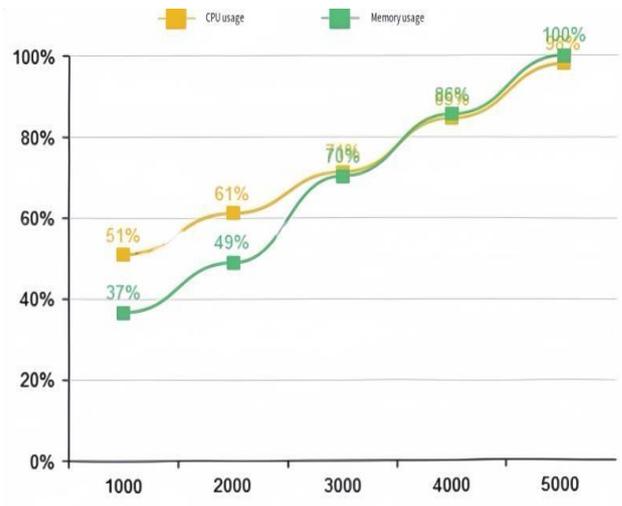

Figure 7. CPU and memory utilization of enterprise big data automatic processing platform based on microservices

In order to verify the security of the platform, testing covered multiple aspects such as data encryption transmission, user access permission restrictions, and data backup. At the same time, the platform has authenticated the identities of different user roles and set corresponding access permissions to ensure that only authorized users can access sensitive data, thereby preventing malicious users from tampering or stealing information. The platform also sets session expiration dates to ensure that sessions that have not been used for a long time can be automatically logged out, and prompts users to log in again when the session expires, further enhancing security. In addition, the platform regularly conducts vulnerability scans and security audits to identify and fix potential security vulnerabilities, thereby enhancing the platform's security and stability. Adaptability testing was conducted on mainstream browsers such as Chrome, IE, and Firefox to confirm that the platform can run normally in all browsers without compatibility issues. In usability testing, users were able to easily find the required features and operate them smoothly, further verifying the platform's friendliness and ease of use.

### V. CONCLUSION

By conducting a detailed analysis of the platform's functional requirements and performance indicators, a use case model was established to help clarify the platform's characteristics and advantages, providing detailed guidance for system design and implementation. On this basis, the application domain driven design (DDD) method divides the platform into multiple microservice modules, such as data collection, parsing, cleaning, and visualization, and completes the interface, data, and service governance design of each module. The verification results indicate that the microservice architecture effectively decouples the system, improves the independence and stability of each service, and ensures efficient communication between services through containerization and Kubernetes cluster deployment, laying a solid foundation for business expansion. In addition, combined with the DSOM microservice scheduling algorithm, the data automatic monitoring and collection process has been redesigned, greatly improving the efficiency and quality of data collection, and solving the speed and quality problems in big data processing. Although the microservice transformation has been completed and its effectiveness has been verified, further optimization is still needed in the future, including containerizing the UI interface of the data collection simulator to improve collection efficiency; Real time monitoring of data parsing and cleaning services to quickly detect anomalies and issue alerts; Combined with artificial intelligence technology, further enhance data analysis capabilities and achieve the intelligent goals of the platform.